**Transmission of droplet-conveyed infectious agents such as SARS-CoV-2 by speech and vocal exercises during speech therapy: preliminary experiment concerning airflow velocity.**

**Authors**:

Antoine Giovanni[1,2*], Thomas Radulesco[2,3*], Gilles Bouchet[3], Alexia Mattei[1,2], Joana Révis[1,2], Estelle Bogdanski[1,2], Justin Michel[2,3]

**Affiliations:**

1. Aix Marseille Univ, CNRS, LPL, Aix-en-Provence, France

2. Aix Marseille Univ, APHM, Conception University Hospital, ENT-HNS Department, Marseille, France

3. Aix Marseille Univ, CNRS, IUSTI, Marseille, France

*\* Pr Giovanni and Dr Radulesco equally contributed to the paper and are co-first authors*

**Corresponding author:**

Antoine Giovanni

antoine.giovanni@ap-hm.fr

Service ORL, CHU Conception,147 boulevard Baille, 13385 Marseille Cedex 5

Tel: 04 91 43 55 25

Fax: 04 91 43 54 17

## Abstract

**Purpose**

Infectious agents, such as SARS-CoV-2, can be carried by droplets expelled during breathing. The spatial dissemination of droplets varies according to their initial velocity. After a short literature review, our goal was to determine the velocity of the exhaled air during vocal exercises.

**Methods**

A propylene glycol cloud produced by 2 e-cigarettes' users allowed visualization of the exhaled air emitted during vocal exercises. Airflow velocities were measured during the first 200 milliseconds of a long exhalation, a sustained vowel /a/ and varied vocal exercises. For the long exhalation and the sustained vowel /a/, the decrease of airflow velocity was measured until 3 seconds. Results were compared with a Computational Fluid Dynamics (CFD) study using boundary conditions consistent with our experimental study.

**Results**

Regarding the production of vowels, higher velocities were found in loud and whispered voices than in normal voice. Voiced consonants like /ʒ/ or /v/ generated higher velocities than vowels. Some voiceless consonants e.g. /t/ generated high velocities, but long exhalation had the highest velocities. Semi-occluded vocal tract exercises generated faster airflow velocities than loud speech, with a decreased velocity during voicing. The initial velocity quickly decreased as was shown during a long exhalation or a sustained vowel /a/. Velocities were consistent with CFD data.

**Conclusion**

Initial velocity of the exhaled air is a key factor influencing droplets trajectory. Our study revealed that vocal exercises produce a slower airflow than long exhalation. Speech therapy should therefore not be associated with an increased risk of contamination when implementing standard recommendations.

## Declarations

**Funding source:** The authors declare no funding source.

**Conflict of interest/Competing interests:** The authors declare no conflict of interest.

**Ethics approval:** Not applicable

**Consent to participate:** Informed consent was obtained from all individual participants included in the study.

**Consent for publication:** Consent for publication was obtained from all individual participants included in the study.

**Availability of data and material:** all data and materials support our published claims and comply with field standards.

**Code availability:** software applications support our published claims and comply with field standards.

**Authors' contributions:** All authors contributed to the study conception and design. Material preparation, data collection and analysis were also performed by all authors. The first draft of the manuscript was written by AG and TR and all authors commented on previous versions of the manuscript. All authors read and approved the final manuscript.

**Introduction**

Airborne infectious agents such as SARS-CoV-2, responsible for COVID-19 disease, are transmitted by droplets of different sizes expelled in the exhaled air, especially during violent events such as cough, but also during speech and singing [1,2]. Health workers, e.g. Otolaryngologist, phoniatricians and speech therapists are therefore at high risk of viral exposure during vocal exercises [3]. Associated with ventilated room and physical distancing, masks have shown to reduce the risk of contamination for the wearer [4,5]. Few data concerning loud voicing, singing and/or vocal exercises are available in literature; moreover, these activities do not always allow the patient to wear a mask. A security issue in the field of vocal speech therapy, singing lessons or in choirs arose during the outbreak. There is a lack of knowledge allowing to resume normal activities.

**State of the literature**

The risk of contamination during vocal exercises was highlighted in March 2020 within a choir in North America [6].

Current data from the literature establish that the risk of contamination mainly depends on three factors related to the emitted droplets: (i) the intrinsic contagiousness of droplets, (ii) the type of vocal exercise, (iii) the spatial dissemination of droplets.

*The intrinsic contagiousness of droplets*

The droplets emitted by an individual contain all the mucus elements and, potentially, a positive viral load if the emitter is infected. The contagiousness of biological fluids is proportional to their viral load [2]: large droplets contain more viruses and therefore have a higher contagiousness than small droplets [2]. The probabilities that 50-, 10- and 1-micron droplets express (after dehydration) at least one copy of the virus are respectively 37%, 0.37% and <0.01% [7]. Very small droplets would therefore be less contaminating. However, being able to remain suspended in the air, they were shown to be responsible for SARS-Cov-2 titrations in ambient air for 3 hours [8]. Therefore, the time spent with an infected individual could be another key factor. Otherwise, the great number of small droplets expelled probably lead to significant risk of contamination [9].

*The type of vocal exercise*

It is proven that more bioaerosols are generated during speech than during calm breathing [10,11]. The number of droplets larger than 50-100 microns expelled during speech is estimated between 1 and 50 particles per second while less than 2 particles per second for normal breathing [12,13]. Consonants /p/, /t/, /s/ generate more droplets (1.8/cm$^3$) than vowels (1/cm$^3$) and normal breathing (0.1/cm$^3$) [14,15]. Asadi *et al*. have shown that droplets emission rates increase with the loudness of the voice, regardless of the language used [14,15]. Anfinrud *et al*. [16] also demonstrated that droplets from 20 to 500 microns emitted in loud voices are twice as numerous as in normal voice. Droplets emitted during speech probably partly originate from the vocal folds' collision during phonation: they could derive from microscopic ruptures of vocal folds lubricating mucus [17]. Indeed, the size distribution of droplets differs between vocalization and whisper [9]. One would think that vocal exercises and singing are considered as violent events leading to a greater number of large droplets than in normal voice, although not demonstrated by scientific evidences. Increased breathing amplitudes, articulatory work and the use of semi occluded vocal tract (SOVT) exercises are supposed to generate more droplets [3].

*The spatial dissemination of droplets*

The contamination of an individual depends on his physical exposure to the infectious agent and is therefore closely related to the spatial dissemination of droplets emitted by an infected individual.

The spatial distribution of droplets emitted depends on their sizes and initial velocities. When speaking, large (> 50-100 microns), intermediated-sized (10-50 microns) and small (<10 microns) droplets are emitted [18]:

- Large droplets have ballistic trajectories, without significant evaporation, which directly depends on their initial velocities [19]. They are quickly slowed down by ambient air and fall to the ground in one to two seconds [12]. They are propelled up to 6 meters during sneezing (v=50 m/s), 2 meters during cough (v = 10 m/s) and less than 1 meter during normal breathing (v=1 m/s) [20,21]. These distances may vary depending on the air distribution in rooms (ventilation systems, patient's own movement) [5, 22].

- Intermediate-sized droplets make the risk of contamination significant because they have trajectories more difficult to predict that strongly depends on their initial velocity but also turbulence and/or shears effects [23]. Their speed of fall depends on their dehydration: a 50 microns droplet reduced to 10 microns by dehydration sees its speed of fall decrease from 6.8 cm/s to 0.35 cm/s [7].

- Small droplets are impacted by air movements and have therefore an unpredictable trajectory after the first second. They could be residues of larger dehydrated droplets [5,23]. But they can also correspond to smaller droplets produced at the free edge of the vocal folds because of aerosolization of secretions lubricating the vocal folds [9,17].

To date, the velocity of expired air according to the type of vocal exercise performed during speech therapy is unknow. Specifying these velocities would allow to better predict emission distance of droplets during phonation.

The objective of this study was to determine the velocity of the expired air during vocal exercises, and to compare these data with a Computational Fluid Dynamics (CFD) study.

**Methods**

**Experimental study**

Two subjects (TR, JR) gave their written consent to participate in this study which was carried out in accordance with the Declaration of Helsinki. It was an observational study that did not substantially change the behavior of subjects who are daily e-cigarette users.

A propylene glycol cloud produced by 2 e-cigarettes' users allowed visualization of the exhaled air emitted during vocal exercises (Eleaf® TC40W, 29W, 0 .58 ohms, 4.15V). The room was quiet and closed. The ventilation system (ceiling) lead to air movements inferior than 5 cm/s. Exhaled airflow velocities were measured on video records during the first 200 milliseconds during successive vocal exercises and during 3 seconds for two of them (long exhalation and sustained vowel /a/):

- sustained vowel /a/ (comfortable intensity and loud intensity). We were not able to determine the absolute value of intensity.
- voiceless fricative consonants /f/ and /ʃ/
- voiced fricative consonants /v/ and /ʒ/
- during reading of a French sentence: 'Monsieur Seguin n'avait jamais eu de bonheur avec ses chèvres. Il les perdait toutes de la même façon' (measures on initial /m/ sounds in 'Monsieur' and /t/ sounds in 'toutes' at comfortable intensity, loud intensity and whispered voice.)
- long exhalation after a deep inspiration
- use of a SOVT (straw, 5mm diameter)

Productions were carried out twice most of the times and averaged. Movies were recorded with a Camera (Sony® FDR AX-33) and analyzed using Da Vinci Resolve video editing software (Magic Mirror). Sounds were recorded with the microphone's camera. A distance marker was used to calculate the size of the cloud.

**CFD study**

We compared velocities measured experimentally with a two-dimensional CFD simulation. CFD was performed using Star-CCM + ® software (Siemens®). The boundary conditions were as follows: airflow= 200 mm$^3$/s; initial velocity=1 m/s; humidity=25%; temperature=22 ° C; vertical ambient air displacement=5 cm/s.

**Results**

Expired air velocities varied according to vocal exercises and are reported in Table 1. Concerning the production of vowels, we found higher velocities in loud and whispered voice. The production of voiced consonants like /ʒ/ or /v/ generated higher velocities than vowels. For voiceless consonants, some, e.g. /t/, generated very fast airflows, close to normal breathing. SOVT exercises generated airflows faster than loud speech. Velocities decreased when voicing in the device.

The evolution of airflow velocities during the first 3 seconds of a long exhalation and a sustained /a/ are reported in Figure 1. Evolution of velocities experimentally measured had a profile similar to that of CFD data (Figure 2). The evolution of the droplets cloud's length during the production of a sustained /a/ is reported in Figure 3 and 4 and the corresponding video (Online Resource 1).

**Discussion**

The objective of this study was to evaluate the initial velocity of the exhaled air during speech and vocal exercises. This study completes our knowledge concerning the dissemination of droplets during speech, which have been reported to be more numerous in loud voice [14]. Our results showed great variability in the expired air velocities, ranging from 0.28 to 1.8 m/s depending on the vocal exercise performed. However, these velocities remained lower than velocities reported for violent events such as coughing and sneezing. [20,21].

The use of propylene glycol for visualization of the exhaled air is a relevant method: e-cigarettes deliver 0.3 microns droplets whose permanence in the air is around 11 seconds, this allowing to calculate exhaled air velocities [24]. However, the future of this cloud after this deadline cannot be accurately evaluated. The deep inspiration required using e-cigarettes was comparable to the deep inspiration required in speech therapy (one of the two subjects is a speech therapist), especially before performing a loud voicing or SOVT exercises. This type of inspiration can also be compared to that used in singing.

The variability of measures reported in figure 1 is probably related to the uncertainties of measurements because velocities were calculated on a very short period (200ms). This is a preliminary study carried out during containment without full access to the technical resources (e.g. velocimeters or lasers) which would have allowed accurate analysis. However, our data agreed with the findings of Anfinrud *et al*. [16].

In all vocal exercises, the initial velocity remained inferior to 1 m/sec for vowels and voiced consonants, even in loud voicing. These data perfectly match with literature: a pressure drop is found at glottic level during voicing and there is no proportionality between the airflow and the acoustic power of the emitted sound. [25]. This velocity becomes close to 1 m/s or slightly higher in the production of voiceless consonants, whispered voicing or SOVT, i.e. in productions where airflow is close to normal expiratory airflow.

Our data suggest that the size of the exhaled cloud should remain limited, even in loud voice. As a result, the risk of spatial dissemination of the droplets is not significantly modified compared to long exhalations. Assuming the speaker is COVID-19 positive, high level personal protective equipment are still needed for healthcare workers. However, we do not report an increased risk during loud voicing or vocal exercises compared to standard speech. The usual rules of physical distance, or wearing a mask in cases where this distance cannot be respected or in a closed place, seem valid. More studies should be performed especially regarding choral singing.

**Conclusion**

The size and velocity of expelled droplets are key factors in transmission of infectious agents. Our study revealed that velocities of the expired air produced during vocal exercises are slower than in long exhalation, suggesting no increased risk compared to standard speech. More studies must be carried out to better understand the infectious risk associated with droplets in speech.

**Tables**

|  | Speaker 1 (Female) Average velocity (minimum-maximum) cm/s | Speaker 2 (Male) Average velocity (minimum-maximum) cm/s |
|---|---|---|
| Long exhalation | 180 | 77 (64-88) |
| Exhalation through a 5mm straw (SOVT) | 102 | 82 (72-90) |
| Voicing through a 5mm straw (SOVT) | 80 (75-90) | 82 (72-90) |
|  |  |  |
| /a/ | 38 (25-44) | 28 (14-33) |
| /a/ « louder » | 48 (40-57) | - |
|  |  |  |
| /m/ | 67 (31-84) | 43 (25-63) |
| /m/ louder | 57 (52-64) | 76 (68-87) |
| /m/ whispered | 90 (80-100) | 78 (57-123) |
|  |  |  |
| /t/ | 84 (73-87) | 80 (78-100) |
| /t/ louder | 80 (75-82) | 91 (72-118) |
| /t/ whispered | 105 (97-120) | 100 (76-118) |
|  |  |  |
| /f/ isolated and brief | 132 | 99 (68-144) |
| /f/ sustained | 94 (88-110) | 108 (104-113) |
| /v/ sustained | 79 | 72 |
|  |  |  |
| /ʃ/ isolated and brief | 154 | 81 (72-90) |
| /ʃ/ sustained | 76 (66-92) | 68 (57-79) |
| /ʒ/ sustained | 44 | 77 |

**Table 1** Velocity (cm/s) of airflow during the first 200 milliseconds

**Figures**

**Fig. 1** Evolution of airflow velocities (in cm/s) during the first 3 seconds of two long exhalations for subject n ° 2 (A) and during the first 3 seconds of a sustained vowel /a/ for subject 1

**Fig. 2** Evolution of expiratory airflow velocity during the digital simulation of a defined expiration at 1 m/s in a calm environment. A: Lateral view. The deceleration of speeds during expiration is comparable to the data recorded experimentally; B: Upper view

**Fig. 3** Evolution of the cloud's length during the first 3 seconds of a sustained vowel /a/

**Fig. 4** Visualization of the cloud during a sustained vowel /a/. A: subject 1, 1 s, lateral view; B: subject 1, 5 s, lateral view; C: subject 2, 2 s, superior view

**Online Resources**

Online Resource 1: Visualization of the cloud according to ample expiration and sustained vowel /a/ (subject 2, lateral view)

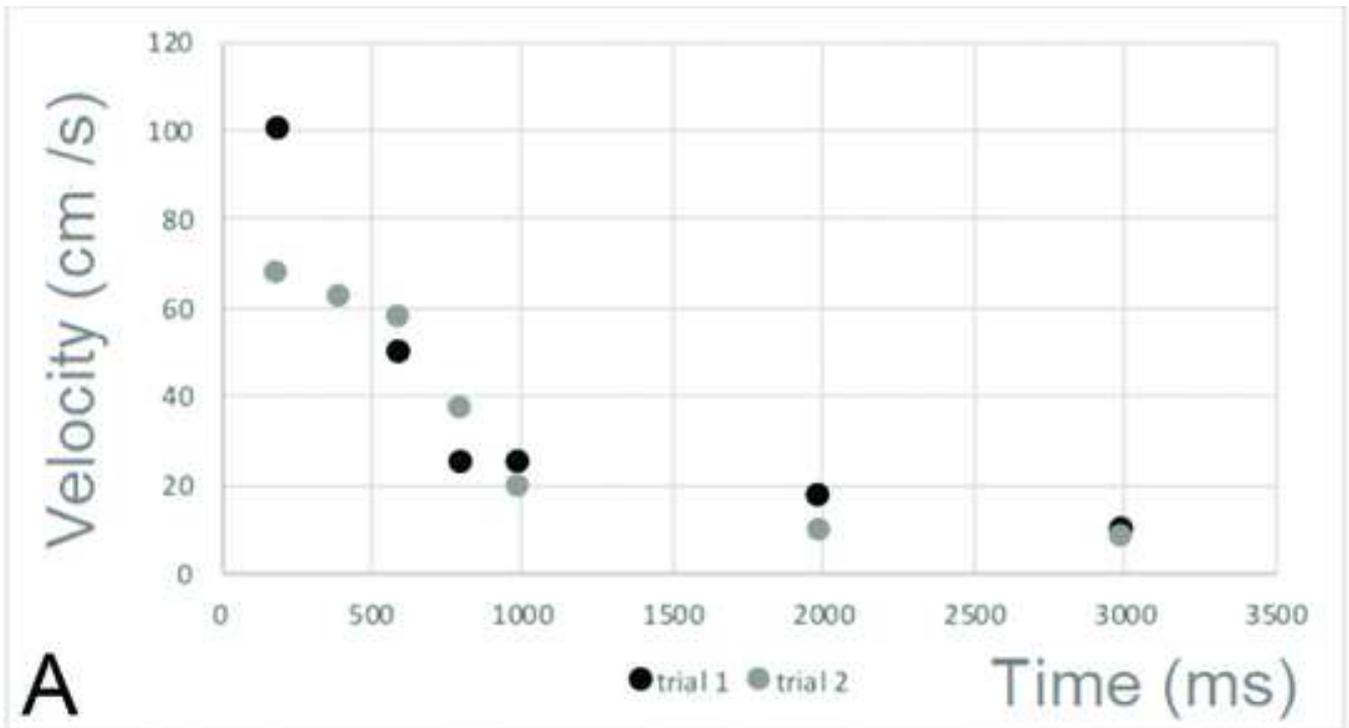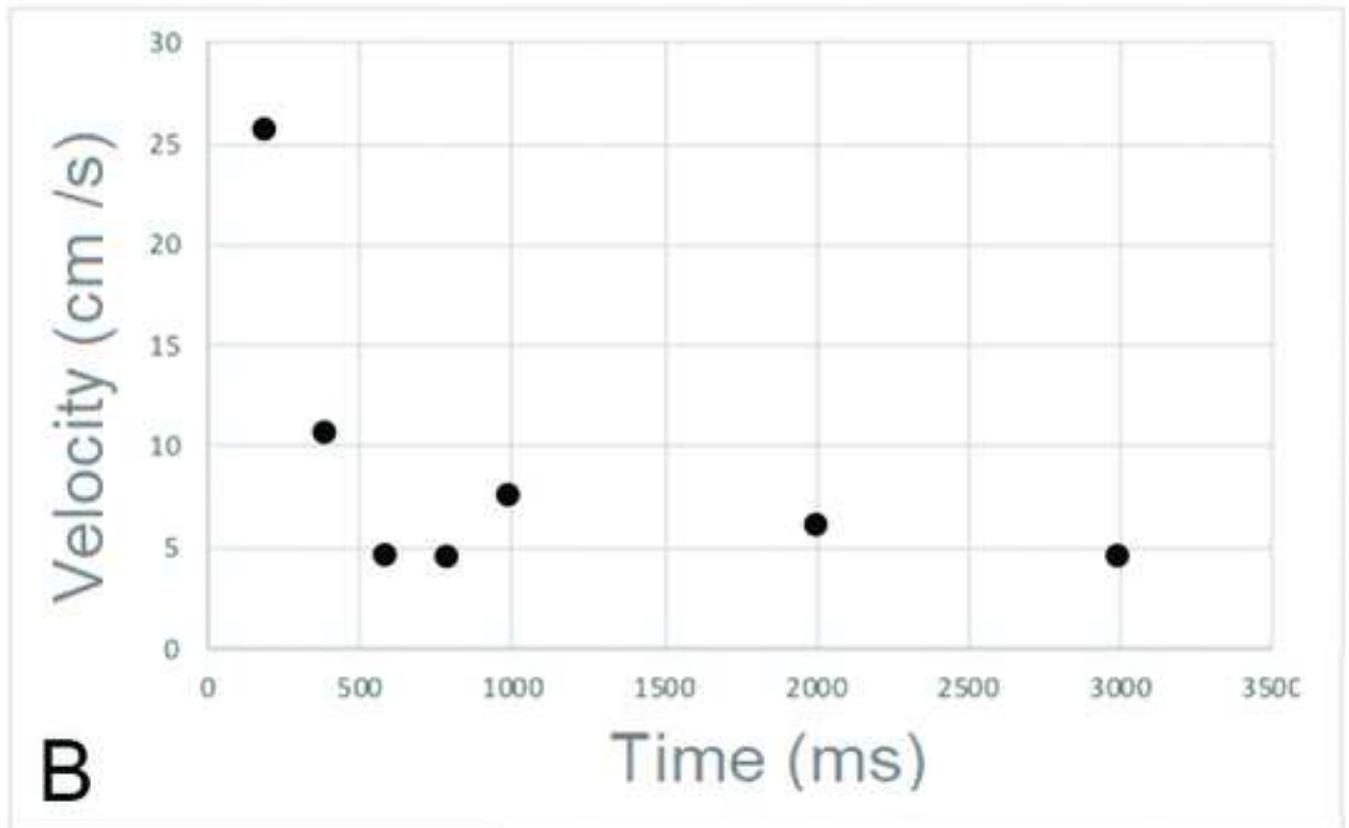

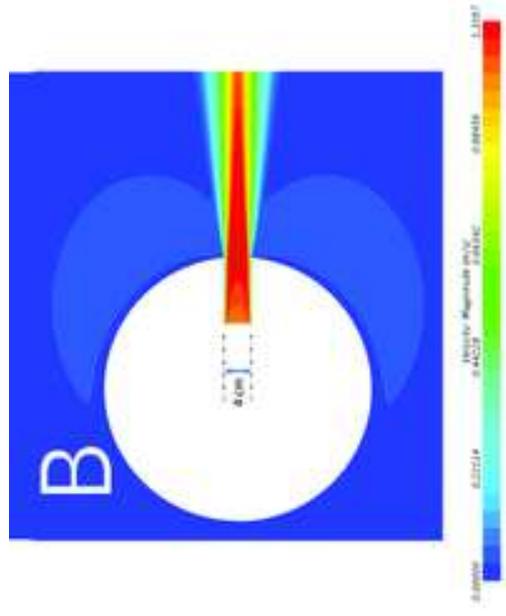

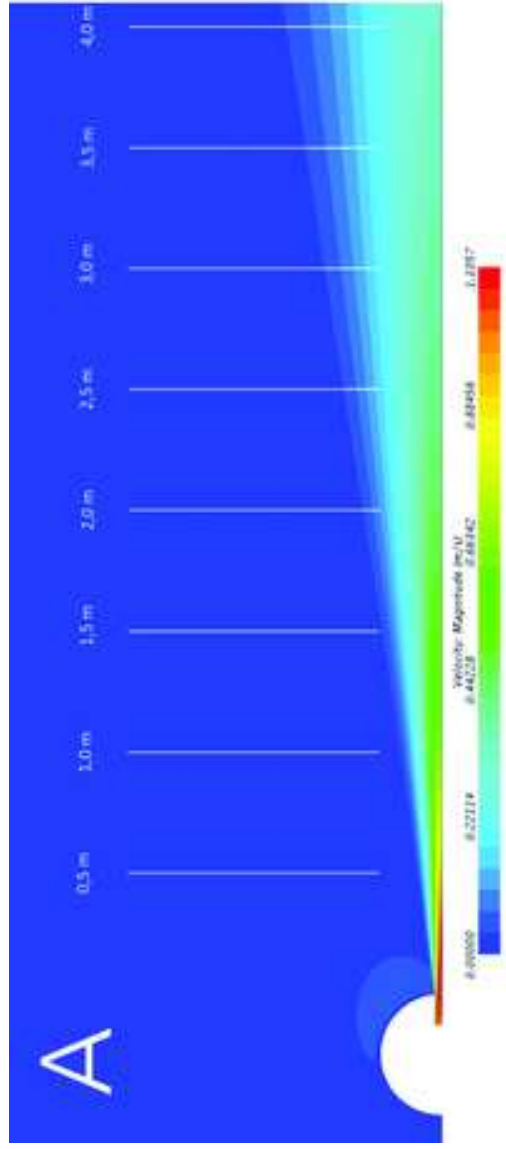

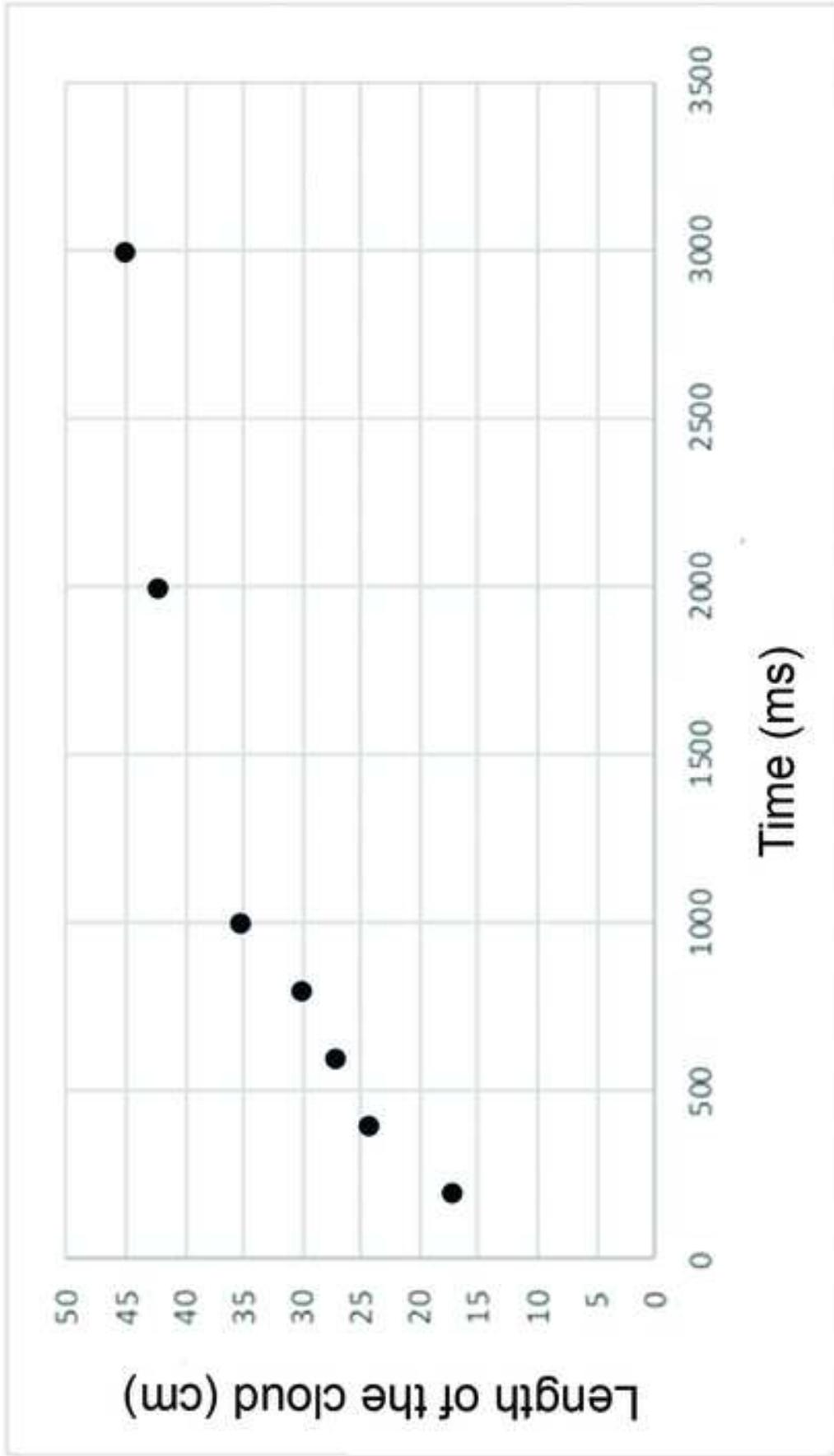

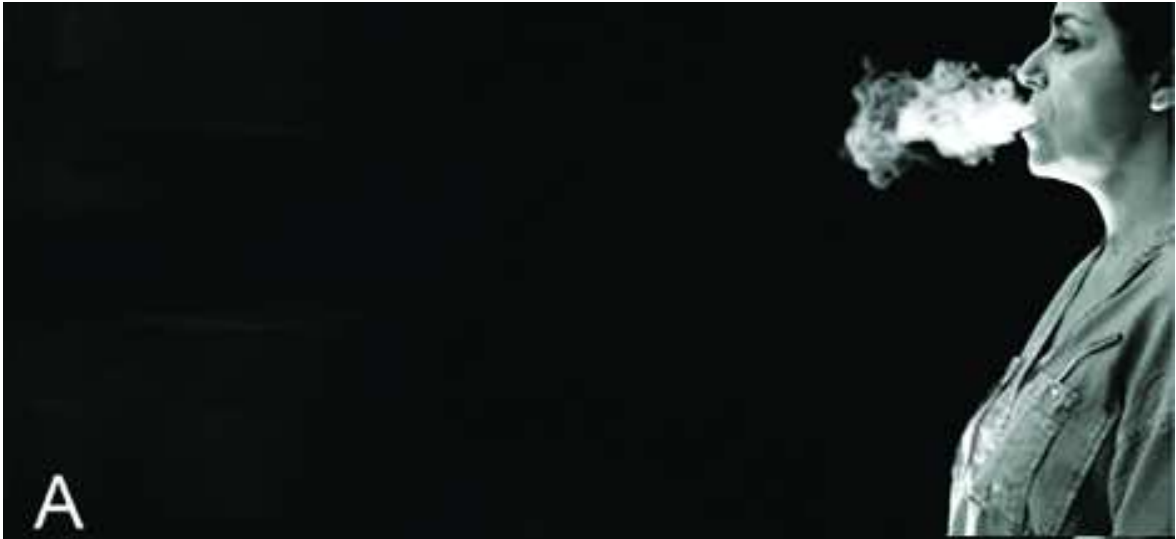
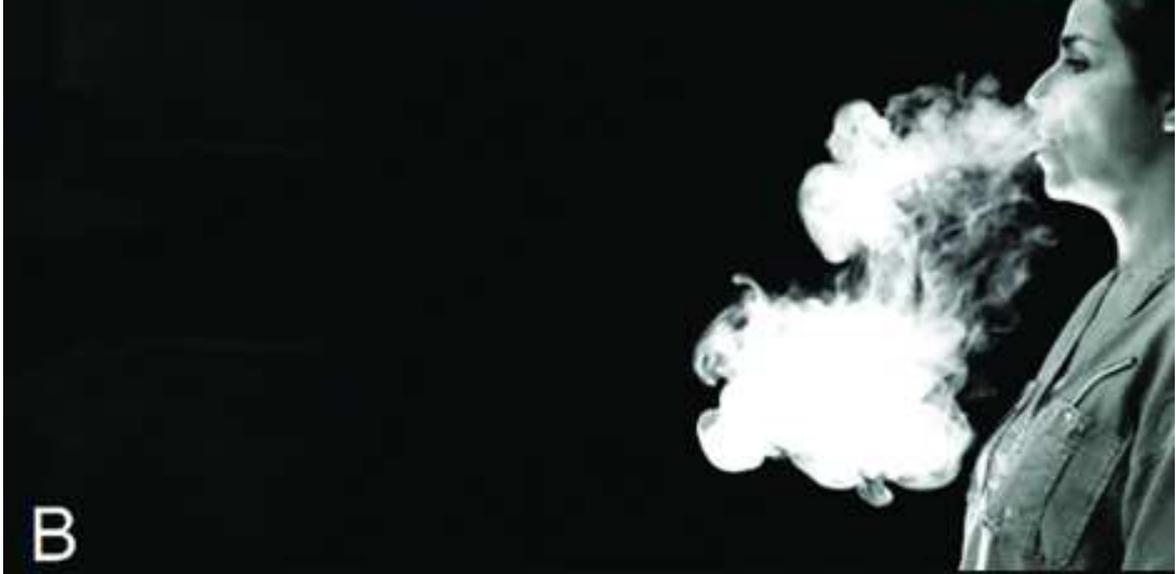
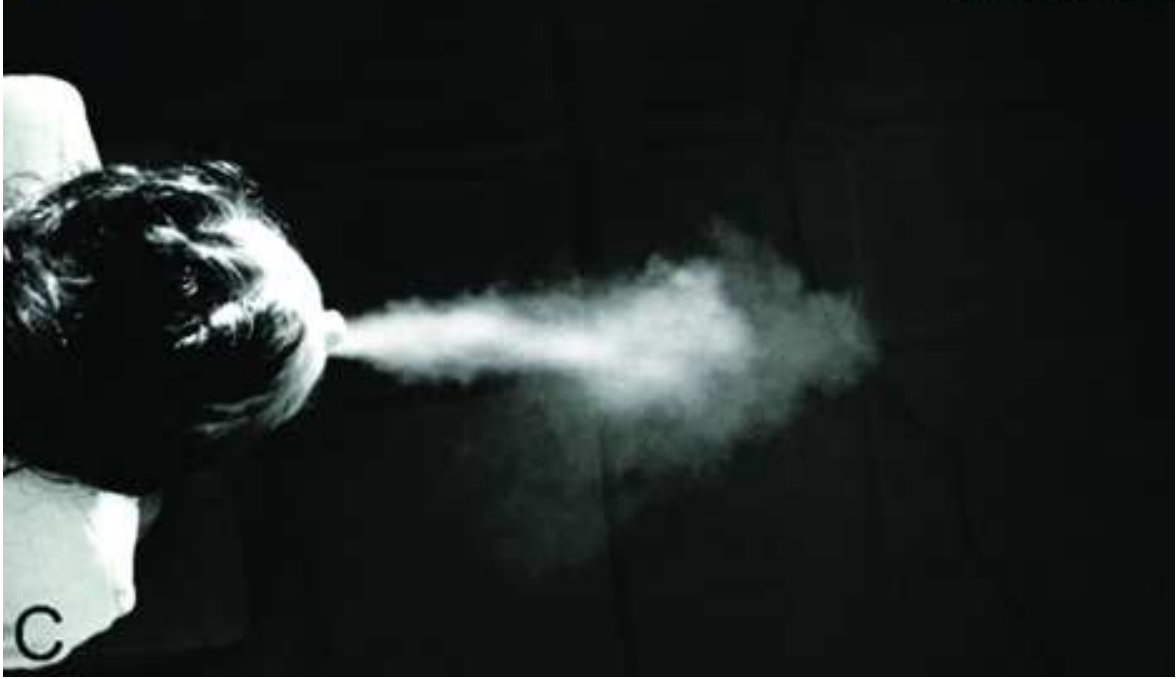